\documentclass[fleqn,usenatbib]{mnras}

\usepackage{newtxtext,newtxmath}

\usepackage{amsmath}

\usepackage[T1]{fontenc}

\DeclareRobustCommand{\VAN}[3]{#2}
\let\VANthebibliography\thebibliography
\def\thebibliography{\DeclareRobustCommand{\VAN}[3]{##3}\VANthebibliography}

\usepackage{graphicx}	
\usepackage{amsmath}	

\title[ML Analysis of Exoplanet Transmission Spectra]{Unsupervised Machine Learning for Exploratory Data Analysis of Exoplanet Transmission Spectra}

\author[K. Matcheva et al.]{
Konstantin T.~Matchev,
Katia Matcheva,\thanks{E-mail: matcheva@ufl.edu}
and Alexander Roman
\\
Physics Department, University of Florida, Gainesville, FL 32611, USA
}

\date{Accepted XXX. Received YYY; in original form ZZZ}

\pubyear{2022}

\begin{document}
\label{firstpage}
\pagerange{\pageref{firstpage}--\pageref{lastpage}}
\maketitle

\begin{abstract}
Transit spectroscopy is a powerful tool to decode the chemical composition of the atmospheres of extrasolar planets. In this paper we focus on {\em unsupervised} techniques for analyzing spectral data from transiting exoplanets. We demonstrate methods for i) cleaning and validating the data, ii) initial exploratory data analysis based on summary statistics (estimates of location and variability), iii) exploring and quantifying the existing correlations in the data, iv) pre-processing and linearly transforming the data to its principal components, v) dimensionality reduction and manifold learning, vi) clustering and anomaly detection, vii) visualization and interpretation of the data. To illustrate the proposed unsupervised methodology, we use a well-known public benchmark data set of synthetic transit spectra. We show that there is a high degree of correlation in the spectral data, which calls for appropriate low-dimensional representations. We explore a number of different techniques for such dimensionality reduction and identify several suitable options in terms of summary statistics, principal components, etc. We uncover interesting structures in the principal component basis, namely, well-defined branches corresponding to different chemical regimes of the underlying atmospheres. We demonstrate that those branches can be successfully recovered with a K-means clustering algorithm in fully unsupervised fashion. We advocate for a three-dimensional representation of the spectroscopic data in terms of the first three principal components, in order to reveal the existing structure in the data and quickly characterize the chemical class of a planet. 
\end{abstract}




\section{Introduction}
\label{sec:introduction}

The exploration of planets beyond the solar system has progressed from a single planet detection to large-scale, ground- and space-based surveys yielding thousands of newly discovered planets and planetary systems\footnote{The Extrasolar Planet Encyclopedia, \url{http://exoplanet.eu}.}. The scientific interest has now migrated from individual planet detection to planetary characterization in search for presence of different chemical compounds with the tantalizing goal to find planets with habitable environments and to detect signatures of biological life. Infrared spectroscopy of planetary atmospheres during transit events is a powerful tool for probing the chemical composition and the structure of the atmosphere  \citep{Schneider1994,Charbonneau2000}.
Transit spectroscopy targets the detection of commonly present atmospheric gasses that have strong absorption lines in the infrared and leave a distinct imprint on the observed modulation, $M(\lambda)$, of the stellar flux
\begin{equation}
M(\lambda) \equiv \frac{F_O(\lambda) - F_T(\lambda)}{F_O(\lambda)},
\label{eq:Mdef}
\end{equation}
where $F_T(\lambda)$ and $F_O(\lambda)$ are the observed stellar flux during and out of the transit event, respectively. The theoretical aspects of transit spectroscopy have been well developed \citep{Seager2000,Brown2001,Hubbard2001,Burrows2003,Fortney2005,Benneke2012,deWit2013,Griffith2014,Vahidinia2014,Heng2015,Betremieux2017,Heng2017} and have been successfully used to extract information about the atmospheric parameters of numerous transiting exoplanets \citep{Cobb2019,Barstow2020,Kitzmann2020,Harrington2021,Cubillos2021,Blecic2021,Welbanks2021}. A fundamental ingredient in any such analysis is a numerical forward radiative transfer model, which, given a set $\mathcal{S}$ of atmospheric parameters, models the atmospheric transmission of the planet and generates a synthetic spectrum $M(\lambda_i)$ at several different wavelengths $\lambda_i$: 
\begin{equation}
    \mathcal{S} \equiv \{ T,P,n_j,m,\kappa_{cl},g, R_0, R_S\} \longrightarrow \{M(\lambda_i)\}.
\label{eq:forward}    
\end{equation}
The set of inputs $\mathcal{S}$ typically includes temperature $T$, pressure $P$, number density $n_j$ of several\footnote{In this paper the index $i$ denotes individual wavelengths, while the index $j$ labels individual gas constituents.} gas constituents ($j=1,2,\ldots$), mean molecular mass $m$, cloud opacity $\kappa_{cl}$, specific gravity $g$, and observational geometry (reference planet radius $R_0$ and stellar radius $R_S$).

The ultimate goal of exoplanet transit spectroscopy is the inversion of the observed transit spectrum, $\{M(\lambda_i)\}$, in order to retrieve the parameters of the planet and its atmosphere:
\begin{equation}
     \{M(\lambda_i)\} \longrightarrow \mathcal{S} = \{ T,P,n_j,m,\kappa_{cl},g, R_0, R_S\}.
\label{eq:inversion}    
\end{equation}
Practical implementations of (\ref{eq:inversion}) rely on various numerical inversion techniques, which recently increasingly utilize novel statistical and machine learning (ML) methods aimed at improving the accuracy, the precision, and the speed of the performed retrievals \citep{Marquez2018, Zingales2018, Cobb2019, Oreshenko2020, Himes2020, Fisher2020, Guzman2020, Nixon2020, Yip2021, Ardevol2021}. The last aspect is becoming especially relevant in view of the sheer number of exoplanet transits expected to be observed over the next several years.

From the ML point of view, both the forward problem (\ref{eq:forward}) and the inverse problem (\ref{eq:inversion}) are in the realm of supervised learning (multivariate regression), where, given a set of features (inputs), we try to predict a set of targets (outputs). The main difference between (\ref{eq:forward}) and (\ref{eq:inversion}) is the choice of variables to be treated as features and targets. The success of any such supervised learning task crucially depends on the availability of high-quality labelled training data, which in turn requires a reliable simulation of the complex phenomenon of radiative transfer. In other words, the inversion (\ref{eq:inversion}) will be only as good as the quality and accuracy of the forward model, and in this sense, inversion results are model-dependent.

While most of the transit spectroscopy literature which employs ML techniques focuses on {\em supervised learning}, in this paper we approach the inversion problem (\ref{eq:inversion}) from the point of view of {\em unsupervised learning}, i.e., we shall focus entirely on the set of measured wavelength-dependent modulations $\{M(\lambda_i)\}$, with no prior knowledge of (or reference to) the corresponding atmospheric parameters $\mathcal{S}$. Unsupervised ML methods attempt to probe the underlying structure within the data set, uncovering hidden trends, correlations and associations, thus providing a more open-minded exploratory perspective on the data, in anticipation of (and in preparation for) any supervised learning tasks to follow. An exhaustive review of unsupervised learning is beyond the scope of this paper\footnote{For a recent review, see \cite{Verbeeck2020}.}, so here we shall illustrate only the unsupervised learning tasks most relevant for planetary spectroscopy: data wrangling (Section~\ref{sec:dataset}), pre-processing and initial exploratory data analysis based on summary statistics (Section~\ref{sec:preprocessing}), factorization methods like principal component analysis (Section \ref{sec:PCArotation}), manifold learning and dimensionality reduction (Sections \ref{sec:PCAdimred} and \ref{sec:manifold}), grouping methods (Section \ref{sec:grouping}) which attempt to identify similar groups (Section \ref{sec:clustering}) or anomalies and outliers (Section \ref{sec:anomaly}). Our primary motivation for going through the unsupervised learning checklist (before tackling the inversion problem directly) is that the problem at hand is by its very nature unsupervised --- when we measure the transit spectra, nature does not provide us with a set of answers for the associated atmospheric parameters. It is therefore prudent to ask, how much progress one could make before invoking the target information from the forward model simulations (\ref{eq:forward}).

\section{Synthetic Data}
\label{sec:syntheticdata}

In this paper we shall take advantage of the publicly available synthetic benchmark data set of \cite{Marquez2018} which is based on a simple analytical model derived in \cite{Heng2017}. The use of an analytical formula as a surrogate for a full radiative transfer model has both its advantages and disadvantages: 
\begin{itemize}
    \item {\it Advantages:} direct traceable connection between the input atmospheric parameters and the resulting spectral characteristics; good understanding of the degeneracy in the resulting spectra \citep{Griffith2014,Heng2017,Welbanks2019,  Matchev2021analytical}; transparent and reproducible results.  
    \item {\it Disadvantages:} analytical expressions are often based on  simplifying  approximations: isothermal atmosphere, isobaric transit cords, grey cloud absorption, no atmospheric scattering, etc. While analytical expressions are therefore perhaps not ideal for precise quantitative studies, they are nevertheless adequate for our illustration purposes here. If desired, all of our results in this paper can be reproduced with spectral data from a full-blown numerical forward radiative transfer model (\ref{eq:forward}).
\end{itemize}
 
  The data set of \cite{Marquez2018} is a well-known benchmark in the literature, which has already been used in several related studies \citep{Marquez2018,Cobb2019, Welbanks2019,Matchev2021analytical}. Alternative benchmark data sets can be found in \cite{Goyal2019,Goyal2020}.

\subsection{Analytical Model}
\label{sec:model} 

Following \cite{Heng2017}, the amount of stellar flux blocked by the transiting planet is proportional to the effective area of the planetary disc:
\begin{equation}
\pi R_T^2 = \pi (R_0 + h)^2 \approx \pi R_0^2 + 2 \pi R_0 h,
\end{equation}
where $R_T$ is the observed transit radius of the planet, $R_0$ is a reference planetary radius, and $h$ is the equivalent thickness of the atmospheric layer which is assumed to be relatively thin ($h<<R_0$). 

Assuming a constant temperature profile, \cite{Heng2017} derived the following expression for the transit radius
\begin{equation}
R_T(\lambda) = R_0 + h =  R_0  + H
\bigl[\gamma_E + E_1(\tau_0) 
+ \ln (\tau_0 )
\bigr],
\label{eq:fullformula}
\end{equation}
where $H$ is the pressure scale height 
\begin{equation}
H \equiv \frac{k_B T}{m g},
\label{eq:Hdef}
\end{equation}
$k_B$ and $\gamma_E$ are respectively the Boltzmann and the Euler–Mascheroni constants, 
\begin{equation}
\tau_0(\lambda) \equiv \frac{P_0\kappa(\lambda)}{g}\sqrt{2\pi \frac{R_0}{H}}
\label{eq:tau0def}
\end{equation}
is the optical thickness of the atmosphere along the line of sight at the reference radius $R_0$ and
\begin{equation}
E_1(\tau_0) = \int_{\tau_0}^\infty \frac{e^{-t}}{t} dt
\label{eq:E1}
\end{equation}
is the exponential integral of the first order with argument $\tau_0$.

In the large $\tau_0$ limit the $E_1$ term vanishes:
$$
\lim_{\tau_0\to \infty} E_1(\tau_0) = 0
$$
and (\ref{eq:fullformula}) simplifies to
\begin{equation}
R_T(\lambda) = R_0 + H 
\bigl[\gamma_E + \ln(\tau_0)
\bigr].
\label{eq:simpleformula}
\end{equation}

\begin{figure}
\begin{center}
	\includegraphics[width=0.8\columnwidth]{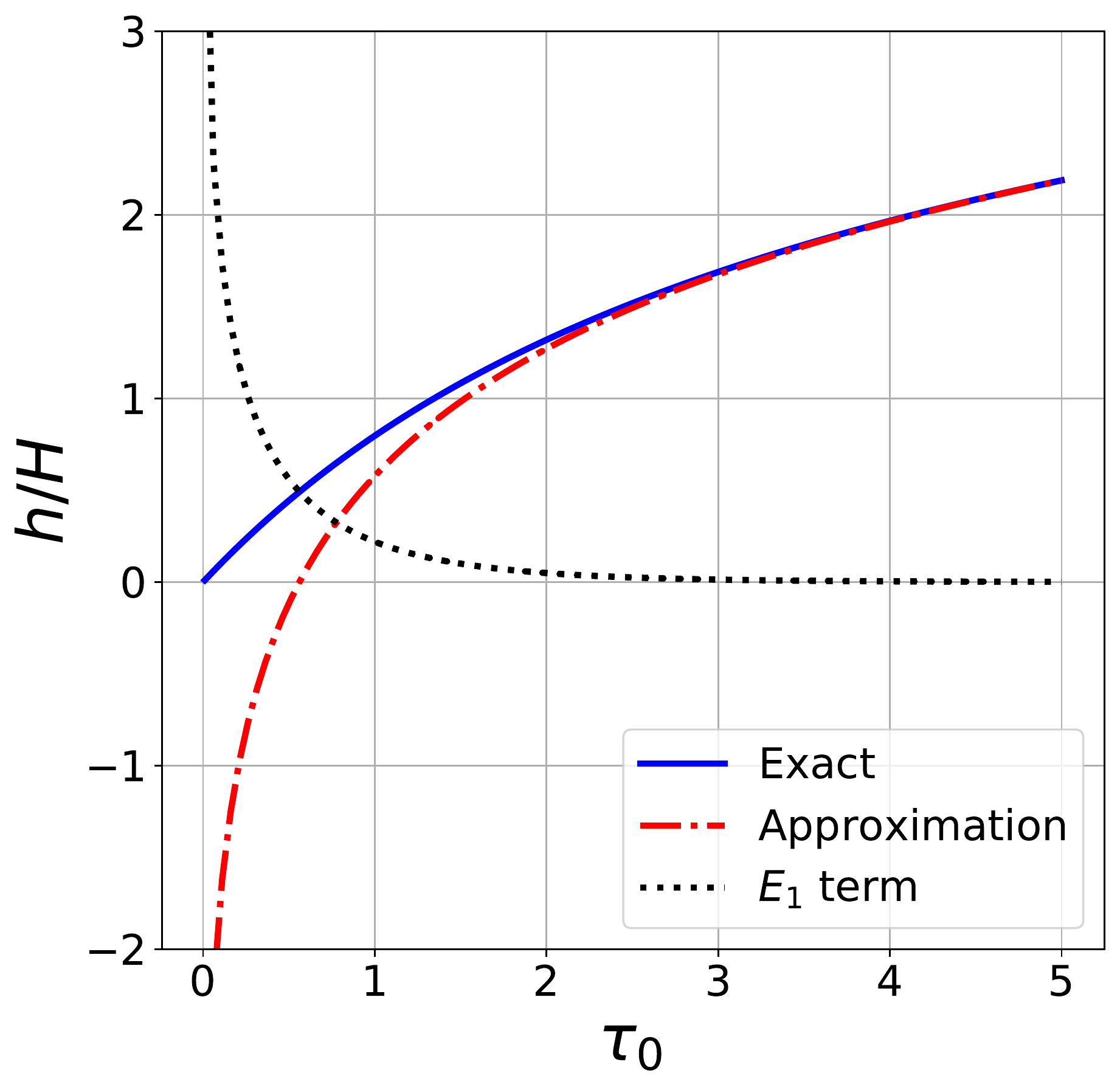}
\end{center}
    \caption{
    The dependence of the equivalent thickness $h$ (measured in units of $H$) on the optical thickness $\tau_0$. The blue solid line gives the exact result (\ref{eq:fullformula}), the red dot-dashed line is the approximation (\ref{eq:simpleformula}), while the black dotted line shows the effect of the $E_1$ term (\ref{eq:E1}).
    }
    \label{fig:E1term}
\end{figure}

The validity of the approximation (\ref{eq:simpleformula}) is illustrated in Figure~\ref{fig:E1term}. We show the dependence of the dimensionless quantity $h/H$ on $\tau_0$ for several cases: the exact result from (\ref{eq:fullformula}) (blue solid line), the approximate result from (\ref{eq:simpleformula}) (red dot-dashed line) and the $E_1$ term (\ref{eq:E1}) alone (black dotted line). As previously pointed out in \cite{Heng2017,Welbanks2019}, for large values of $\tau_0$, the approximation (\ref{eq:simpleformula}) works pretty well, but as $\tau_0$ drops to order $1$ or below, the accuracy of the approximation degrades and one should resort back to the full expression (\ref{eq:fullformula}). These observations are indeed confirmed in Figure~\ref{fig:E1term}.

Note that many of the atmospheric parameters are embedded in the value of $\tau_0$, and specifically in the wavelength-dependent absorption cross-section per unit mass
$\kappa(\lambda)$ which can be related to the atmospheric composition as
\begin{equation}
\kappa = \kappa_{cl} + \sum_{j} \frac{m_j}{m} X_j \kappa_j,
\label{eq:kappadef}
\end{equation}
where $m_j$, $X_j$ and $\kappa_j$ are, respectively, the molecular mass, the volume mixing ratio and the opacity for a given minor gas constituent $j$.

\subsection{The Synthetic Benchmark Data Set}
\label{sec:dataset}

The synthetic benchmark data set of \cite{Marquez2018} consists of 100,000 synthetic Hubble Space Telescope Wide Field Camera 3 (WFC3) transit spectra of hot Jupiters observed at 13 different wavelengths $\lambda_i$ in the range $0.838-1.666\, {\mu}m$. The data set is created by using equations (\ref{eq:simpleformula}), (\ref{eq:Hdef}), (\ref{eq:tau0def}) and (\ref{eq:kappadef}). Five different atmospheric parameters, namely, temperature $T$, H${}_2$O, HCN and NH${}_3$ mixing ratios, and grey cloud opacity $\kappa_{cl}$, were scanned in their respective ranges:
$T\in (500\ {\rm K}, 2900\ {\rm K})$,
$X_{\rm H_2O} \in (10^{-13}, 1)$,
$X_{\rm HCN} \in (10^{-13}, 1)$,
$X_{\rm NH_3} \in (10^{-13}, 1)$,
$\kappa_{cl} \in (10^{-13}, 10^2)$ cm${}^2$g${}^{-1}$.
For each spectrum, the values of the five parameters were randomly generated from a log-uniform (volume mixing ratios and cloud opacities) or uniform (temperature) distribution.
In addition, for temperatures larger than $1500$ K, the ammonia mixing ratio was set to be very small, $X_{\rm NH_3}=10^{-13}$. The remaining atmospheric parameters were held fixed throughout the data set: $R_0 = 1.79 R_J$, $R_S = 1.57 R_{\odot}$, $g=9.77$ m/s$^{2}$, $P_0=10$ bar, $m = 2.4\, m_{\rm amu}$, where $m_{\rm amu}=1.660539040\times 10^{-27}$ kg is the atomic mass unit. The benchmark data set is then provided as a dataframe with 18 entries (columns) per data point (row): the 13 features $\{M(\lambda_i)\}$, $i=1,2,\ldots,13$, followed by the 5 target variables in the form 
\begin{equation}
\left\{T, \log_{10} (X_{\rm H_2O}) , \log_{10} (X_{\rm HCN}) , \log_{10} (X_{\rm NH_3}) , \log_{10} (\kappa_{cl})\right\}. 
\label{eq:targets}
\end{equation}

\begin{figure*}
	\includegraphics[width=0.8\columnwidth]{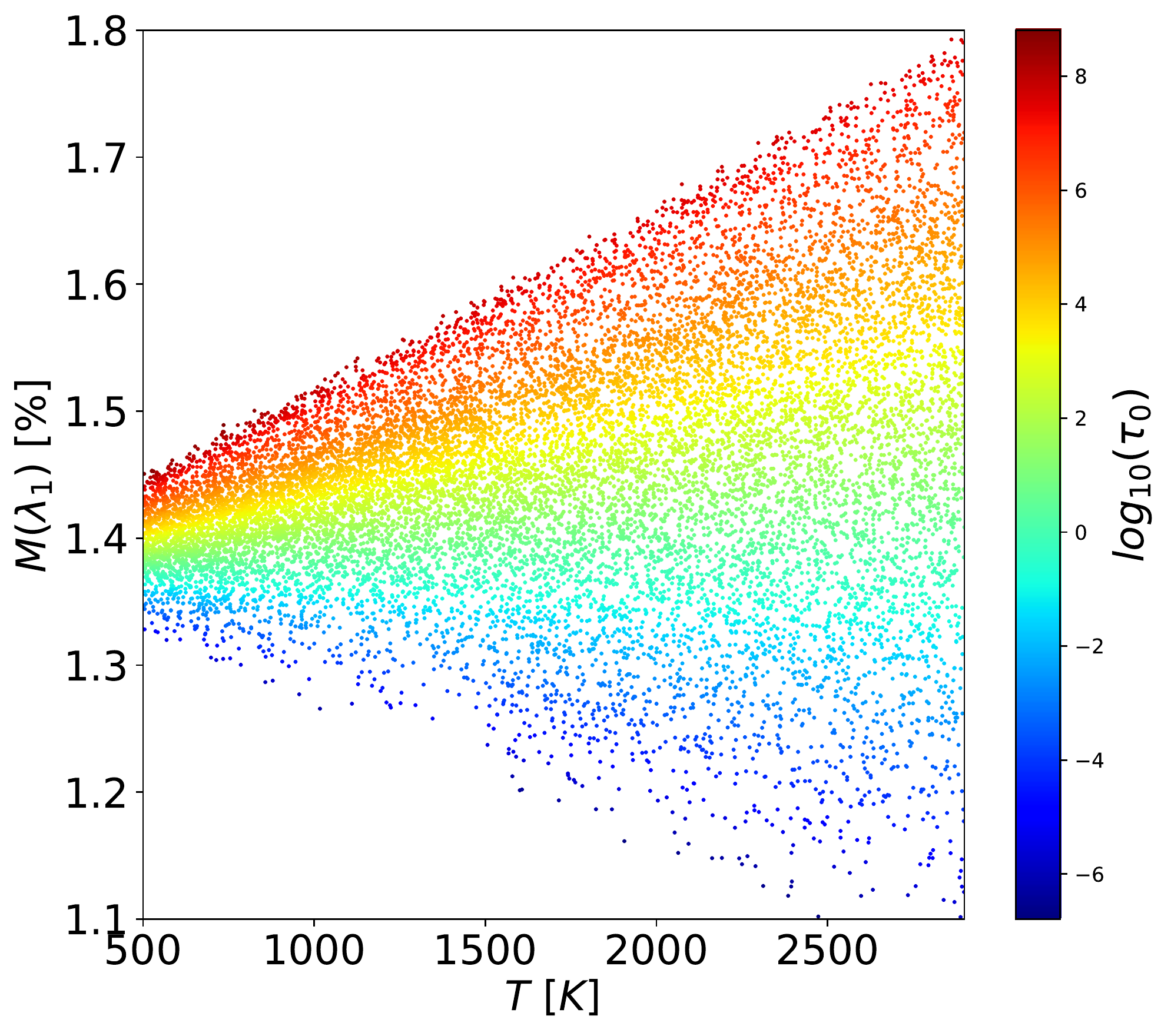}
	\includegraphics[width=0.8\columnwidth]{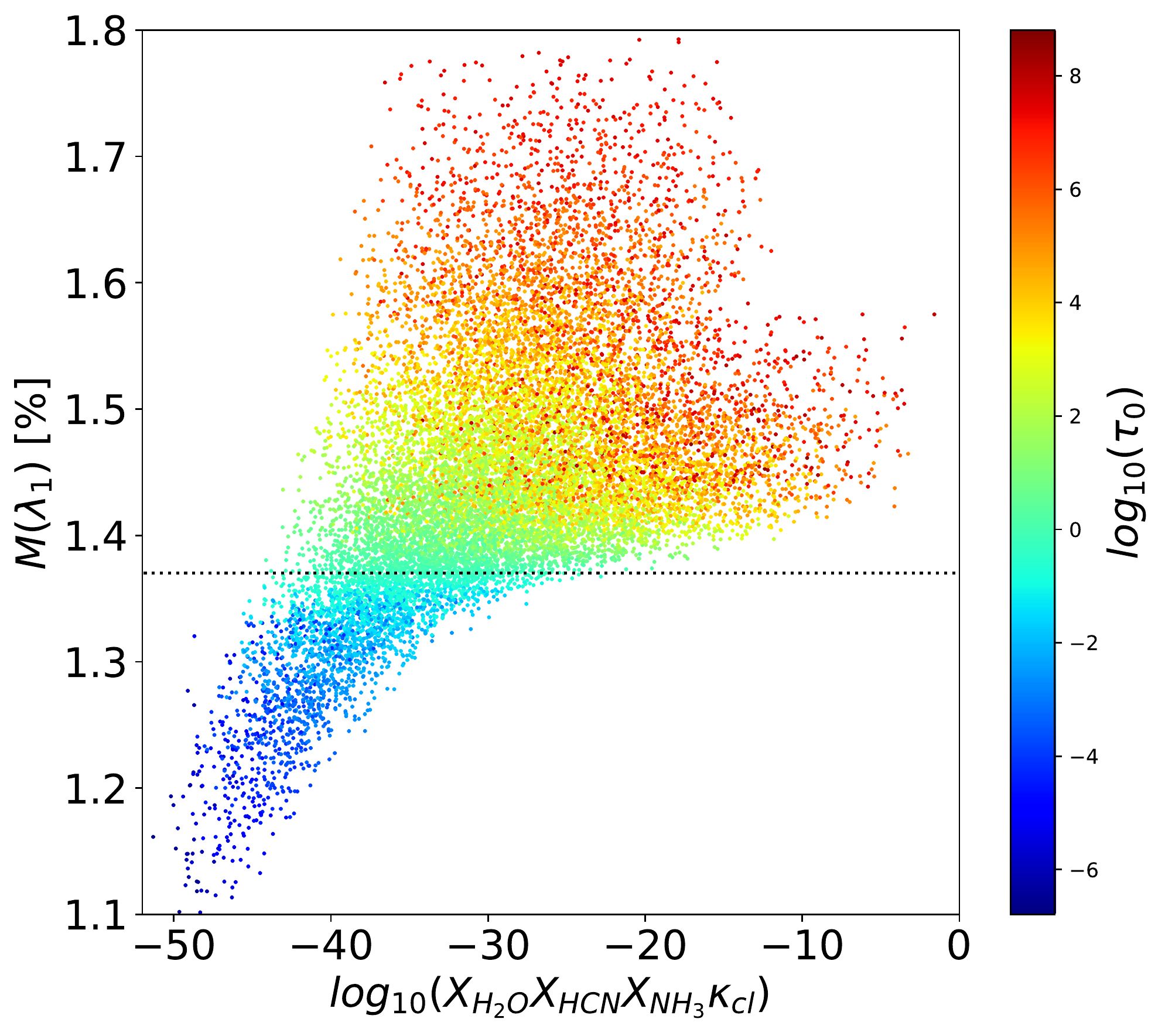}
    \caption{Scatter plots of 15,000 data points from the benchmark data set: $M(\lambda_1)$ versus $T$ (left panel) and $M(\lambda_1)$ versus the variable in (\ref{eq:sumlogs}) which can be equivalently expressed as $\log_{10}(X_{\rm H_2O}X_{\rm HCN}X_{\rm NH_3}\kappa_{cl})$ (right panel). The points are color-coded by the value of $\log_{10}(\tau_0)$ as indicated on the colorbar. The horizontal dotted line in the right panel denotes the constant threshold $M(\lambda_1)=(R_0/R_S)^2\times 100\%$.}
    \label{fig:Mvsparameters}
\end{figure*}

The thirteen features $\{M(\lambda_i)\}$ are functions of the five targets, four of which (the mixing ratios and the cloud opacity) enter only as the combination defined in eq.~(\ref{eq:kappadef}). This implies that the five degrees of freedom are reduced effectively to only two: the temperature T and the total opacity $\kappa$. One can reach similar conclusions on the basis of dimensional analysis as in \cite{Matchev2021analytical}. As an illustration, in Figure~\ref{fig:Mvsparameters} we show scatter plots of one representative feature ($M(\lambda_1)$ for $\lambda_1=0.867\ \mu{\rm m}$) versus the two relevant degrees of freedom: the temperature $T$ (left panel) and a convenient naive proxy for the total atmospheric opacity in eq.~(\ref{eq:kappadef}),
\begin{equation}
\log_{10} (X_{\rm H_2O}) + \log_{10} (X_{\rm HCN}) + \log_{10} (X_{\rm NH_3}) + \log_{10} (\kappa_{cl}). 
\label{eq:sumlogs}
\end{equation}
The points in those plots are color-coded by the value of $\log_{10}(\tau_0)$ as indicated on the colorbars.

Figure~\ref{fig:Mvsparameters} shows the expected behavior of $M(\lambda_1)$: as the temperature and gas concentrations increase, so does the optical thickness $\tau_0$ and accordingly $M(\lambda_1)$ as well. However, we also observe an unusual behavior in the right panel, where at very low concentrations $M(\lambda_1)$ drops below the minimum expected threshold of $(R_0/R_S)^2\times 100\%$ marked with the horizontal dotted line. As noted in \cite{Welbanks2019}, this can be traced to the breakdown of the approximation (\ref{eq:simpleformula}) at low $\tau_0$ (as already illustrated in Figure~\ref{eq:E1}), as well as to the absence of collisionally induced absorption (CIA) in the calculation of the overall opacity $\kappa$. For this reason, we shall exclude any points with $\tau_0<1$ from further considerations below. This still leaves us with more than 80,000 of the original 100,000 spectra in the data set.

Figures~\ref{fig:E1term} and \ref{fig:Mvsparameters} are the only two figures in this paper in which we plot information contained in the target labels (\ref{eq:targets}). From now on we shall completely ignore the target information and in true unsupervised fashion treat the data as unlabelled, focusing exclusively on the 13 features $\{M(\lambda_i)\}$.

\section{Initial Exploratory Data Analysis}
\label{sec:preprocessing}

The first step in any data science project is to explore the data. In this section we perform a preliminary data exploration of our features using standard statistics techniques.

\subsection{Feature Correlations}

For each of the 80,000 sample atmospheres there are 13 individual measurements, henceforth referred to as features, that correspond to the minimum stellar flux during transit at different wavelengths. The immediate question is how much independent information is contained in these 13 measurements. The standard way to answer this question is to investigate the correlations among the individual measurements.

\begin{figure}
\begin{center}
	\includegraphics[width=0.9\columnwidth]{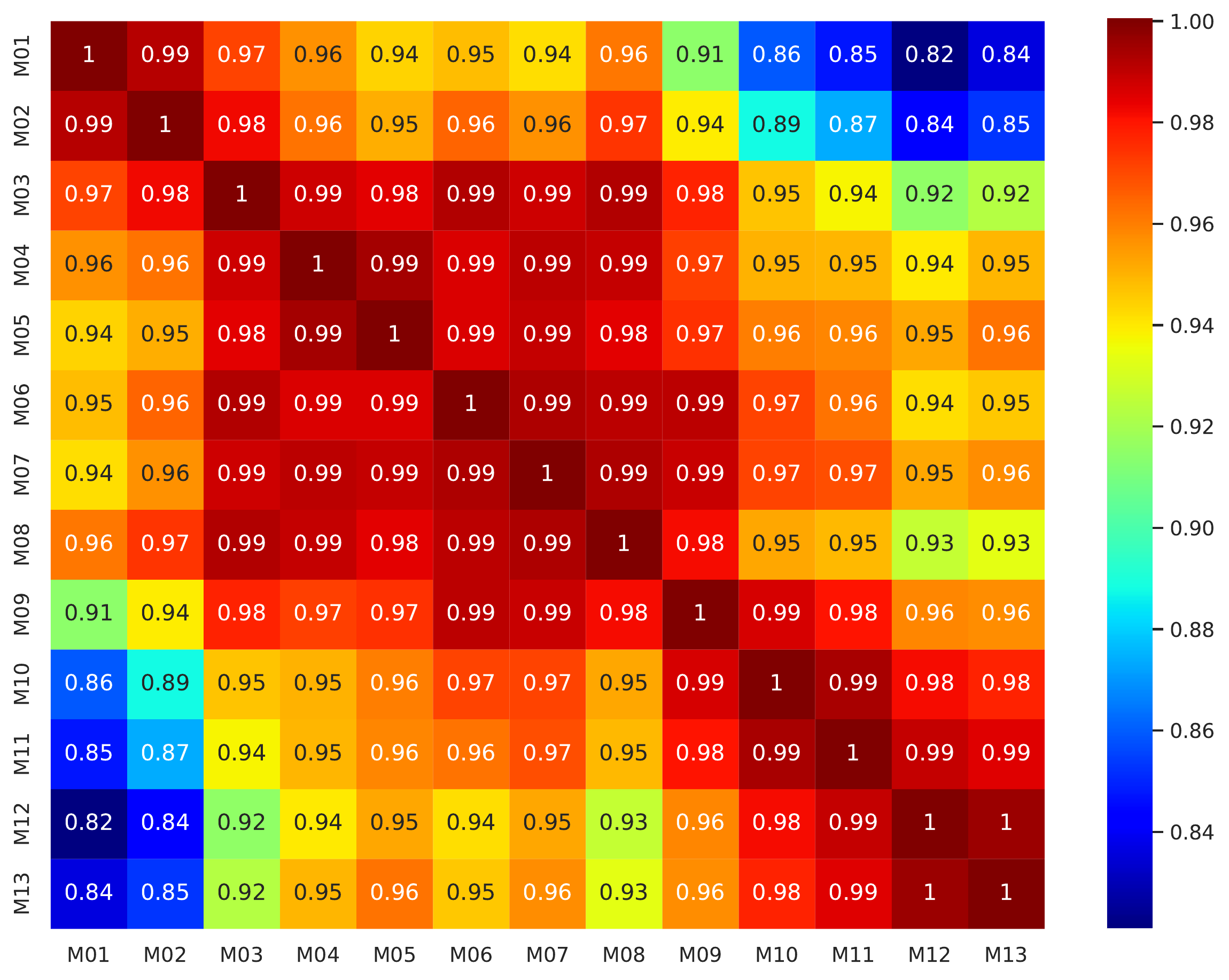}
\end{center}	
    \caption{Correlation matrix of the 13 features $\{M(\lambda_i)\}$, $i=1,\ldots,13$.}
    \label{fig:correlations}
\end{figure}

One way to visualize the correlations among the features in a data set is to look at the scatter plot matrix which, as the name suggests, consists of all possible scatter plots of one feature against another. It turns out that in our case the features $\{M(\lambda_i)\}$ are very highly correlated. Thus, instead of boring the reader with $13\times (13-1)/2=78$ plots of essentially diagonal distributions, in Figure~\ref{fig:correlations} we choose to simply show the correlation matrix which lists the Pearson correlation coefficient between any two features in the data set. This also provides a quantitative measure of the correlations --- the correlation coefficient is listed in each individual box. Figure~\ref{fig:correlations} confirms that the features are indeed very correlated, since the coefficients are very close to 1. We also notice that measurements which are closer in wavelength are more correlated than measurements taken at more distant wavelengths. This motivates expanding the wavelength range of planned measurements. In other words, the sheer increase in the number of wavelengths at which observations are done, does not guarantee adding much new information --- to maximize the information gain, one needs to design observations at wavelengths which are further apart.

\subsection{Summary Statistics of Transit Spectra}
\label{sec:statistics}

The high level of correlations exhibited in Figure~\ref{fig:correlations} implies that the same amount of information can be represented with much fewer variables, which in turn would greatly simplify any subsequent spectral inversions. One of our main goals in this paper will be to find exactly such reduced-dimensionality representations.

Next we shall look at the two most common summary statistics which are applicable to any data: the mean $\mu$ of each spectrum
\begin{equation}
\mu (M) \equiv \frac{1}{13}\, \sum_{i=1}^{13} M(\lambda_i)
\label{eq:mudef}
\end{equation}
and the standard deviation $\sigma$ 
\begin{equation}
\sigma (M) \equiv 
\left[ \frac{1}{13}\, \sum_{i=1}^{13} \biggl(M(\lambda_i) - \mu(M)\biggr)^2 \right]^{1/2}.
\label{eq:sigmadef}
\end{equation}
In other words, each sample spectrum is additionally characterized by these two parameters: $\mu(M)$ and $\sigma(M)$, which in statistics are the standard measures of location and variability, respectively. In our case, $\mu(M)$ represents the typical value of $M(\lambda_i)$,  while $\sigma(M)$ captures the variability of the spectra over the probed range of wavelengths.

\begin{figure*}
	\includegraphics[width=2\columnwidth]{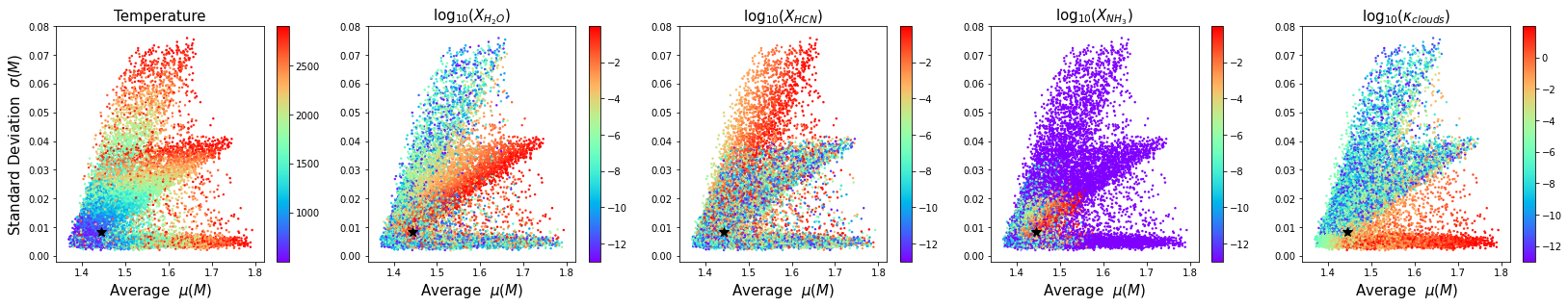}
    \caption{Scatter plots of 15,000 data points: the average $\mu(M)$ (plotted on the $x$-axis) versus the standard deviation $\sigma(M)$ (plotted on the $y$-axis). In each panel, the points are color-coded by the value of one of the five target variables indicated on the top. The black $\star$ symbol marks the location of the hot gas-giant exoplanet WASP-12b. }
    \label{fig:pentaptych}
\end{figure*}

The pentaptych in Figure~\ref{fig:pentaptych} shows scatter plots of 15,000 sample spectra in the $(\mu, \sigma)$ plane. These simple plots already exhibit a very interesting structure: the bulk of the points are clustered in the lower left portion of the point distribution, around $\mu\sim 1.4$ and $\sigma\sim 0.01$. This was to be expected, since the sampling ranges for the three mixing ratios and the cloud opacity extend all the way down to $10^{-13}$, which results in a large number of synthetic atmospheres with very little absorption. 

More interestingly, however, Figure~\ref{fig:pentaptych} reveals three distinct branches extending away from the bulk population. In order to understand the physics behind each branch, we color-coded the points in the figure by the value of one of the five target variables as indicated on the top of each panel. Each branch emerges only at high temperatures, as evidenced in the first panel in Figure~\ref{fig:pentaptych}. Each branch is generated by a specific constituent which is {\em solely} responsible for the increased opacity:
\begin{itemize}
    \item {\em The cloud branch.} As shown in the fifth panel of Figure~\ref{fig:pentaptych}, points in the (horizontal) cloud branch are associated with high values of $\kappa_{cl}$ and relatively low gas absorption (small $X_j$'s). Since the clouds are gray, they contribute to the opacity equally at different wavelengths, which results in a ``flat" spectrum with a low variability with respect to the wavelength.
    \item {\em The {\rm H$_2$O} branch.} As shown in the second panel of Figure~\ref{fig:pentaptych}, points in the (middle) water vapor branch have high values of water concentration and relatively low concentrations for the other absorbers (as evidenced from the remaining panels). Since the absorption due to water vapor is wavelength-dependent, its presence changes both the average and the standard deviation, which results in a distinct slope of the branch in the $(\mu,\sigma)$ plane.
    \item {\em The {\rm HCN} branch.} Finally, points with high hydrogen cyanide abundance and relatively low concentrations for the other absorbers form the third, steepest, branch highlighted in the third panel of Figure~\ref{fig:pentaptych}. The steep slope of the branch in the $(\mu,\sigma)$ plane indicates a large variability in the associated spectrum.
\end{itemize}
In addition, as seen in the fourth panel of Figure~\ref{fig:pentaptych}, there is also an ammonia (NH$_3$) branch which is beginning to emerge, but it terminates early due to the lack of ammonia at high temperatures in the data set. Note that points for which two or more constituents have high concentrations, fall in the space between the branches described above.

The $(\mu,\sigma)$ representation\footnote{For completeness, we also considered other measures of location (the median) and variability (the range and the average slope), but the results were very similar to Figure~\ref{fig:pentaptych}.} depicted in Figure~\ref{fig:pentaptych} is an important result of this paper --- it clearly shows well-defined classes of different spectral behavior which can be used for a ``quick and dirty" estimate of a planet's atmospheric composition. Newly measured transit spectra $\{M(\lambda_i)\}$ of hot gas-giant exoplanets can be plotted in this representation and quickly classified based on $\sigma(M)$ and $\mu(M)$ alone. If a spectrum falls squarely on one of the extended branches, one can immediately deduce the dominant absorber. Alternatively, in the case when the spectrum ends up in the ``bulk" of the distribution, one can rule out both high temperatures and high concentrations of gas absorbers and clouds. As a specific example, the black $\star$ symbol in Figure~\ref{fig:pentaptych} marks the location of planet WASP-12b, revealing that it is a rather nondescript planet, far from any extremes. This qualitative observation is consistent with the result from the quantitative analysis in \cite{Marquez2018}.

\section{Principal Component Analysis}
\label{sec:PCA}

\subsection{Features in the PCA basis}
\label{sec:PCArotation}

The correlation analysis in Figure~\ref{fig:correlations} suggests that the spectral data allows an equivalent lower-dimensional representation. In Figure~\ref{fig:pentaptych} we already identified one such low-dimensional representation in terms of the summary statistics $\mu$ and $\sigma$. But was this the optimal choice? The standard approach to tackling this question is to perform a principal component analysis (PCA) of the data and inspect the explained variance ratio of each component (for an introduction to PCA, see \cite{Jolliffe2002}).

\begin{figure}
\begin{center}
	\includegraphics[width=0.8\columnwidth]{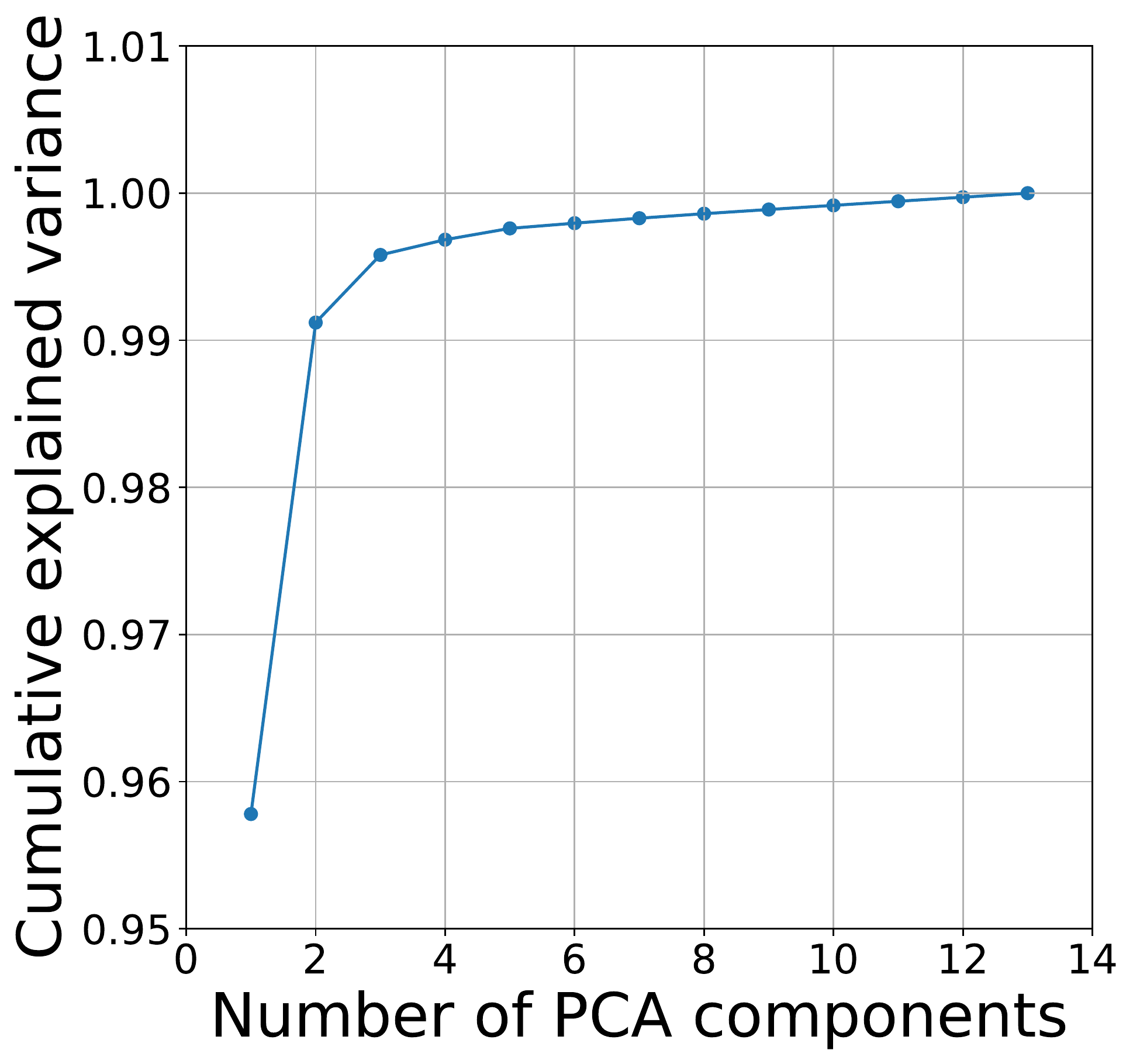}
\end{center}
    \caption{The cumulative explained variance ratio as a function of the number of included PCA components.}
    \label{fig:EV}
\end{figure}

We use the PCA tools available in {\tt scikit-learn} \citep{scikit-learn} to linearly transform our original features $\{M(\lambda_i)\}$ to an equivalent, but uncorrelated, 13-component PCA basis, in which the components are ordered according to their corresponding variance within the sample:\footnote{As eq.~(\ref{eq:PCAdef}) implies, we choose to center but not scale the features since the shapes of their distributions are already very similar.}
\begin{equation}
M(\lambda_i), (i=1,..,13) ~\to~ \sum_{i=1}^{13} \alpha_i^{(k)} \left[M(\lambda_i)-\mu(M)\right], (k=1,..,13). 
\label{eq:PCAdef}
\end{equation}
Figure~\ref{fig:EV} shows the cumulative explained variance ratio as a function of the number of included PCA components. We see that the first principal component alone already accounts for as much as 95.8\% of the variance in the data. Incorporating just one more PCA component brings up the total explained variance to over 99\%, and the information gain beyond the first three PCA components is very minimal, as was also discussed in \cite{Hayes2020}.

\subsection{PCA as dimensionality reduction}
\label{sec:PCAdimred}

\begin{figure}
\begin{center}
	\includegraphics[width=0.48\columnwidth]{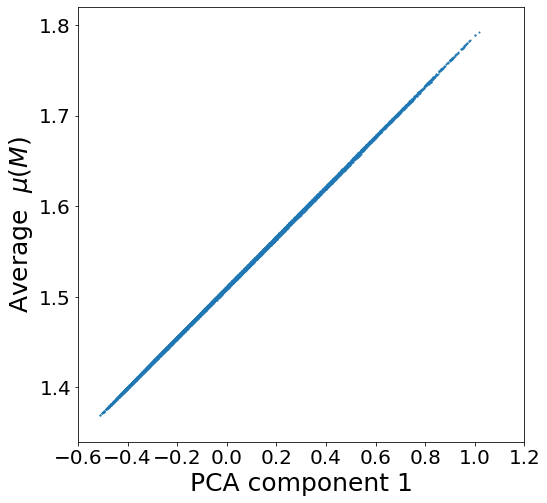}
	\includegraphics[width=0.48\columnwidth]{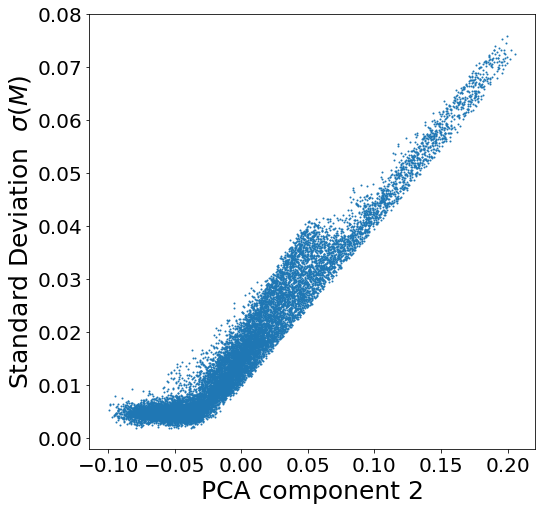}
\end{center}
    \caption{Correlation scatter plots (15,000 points) of the first two PCA components against the summary statistics parameters $\mu(M)$ and $\sigma(M)$, respectively.}
    \label{fig:PCAvsStats}
\end{figure}

Figure~\ref{fig:EV} suggests that we can successfully use just the first few PCA components themselves to represent the data. The numerical values of the coefficients $\alpha_i^{(k)}$ for the first three PCA components ($k=1,2,3$) are listed in Table~\ref{tab:PCA}. Upon inspection, we recognize that the weights with which the individual $M(\lambda_i)$ enter the first PCA component, are almost equal. This implies that the first PCA component is closely related to the average value ($\mu(M)$) already discussed before. This  correlation is, in fact, almost perfect, as illustrated by the scatter plot with 15,000 points in the left panel of Figure~\ref{fig:PCAvsStats}. By a similar argument we arrive at the conclusion that the second principal component reflects the overall slope of the spectrum $M(\lambda)$, which in turn is related to the standard deviation $\sigma(M)$. This is illustrated in the right panel of Figure~\ref{fig:PCAvsStats}, where an observant reader might even spot the three-branch structure discussed in Section~\ref{sec:statistics}.

\begin{table}
 \caption{The coefficients $\alpha^{(k)}_i$ appearing in the linear combination (\ref{eq:PCAdef}) for the first three PCA components ($k=1,2,3$). }
 \label{tab:PCA}
 \begin{tabular}{lrrr}
  \hline
$i$  & $k=1$ & $k=2$ & $k=3$  \\  \hline
 1 & 0.25665 & -0.49546 & 0.18981 \\
 2 & 0.25951 & -0.43182 & -0.12505 \\
 3 & 0.26775 & -0.20320 & -0.06718 \\
 4 & 0.26843 & -0.10890 & 0.34669 \\
 5 & 0.27271 & -0.04528 & 0.37310 \\
 6 & 0.27616 & -0.08210 & -0.19907 \\
 7 & 0.27606 & -0.04170 & -0.02248 \\
 8 & 0.27287 & -0.15565 & -0.05916 \\
 9 & 0.28385 & 0.04698 & -0.44404 \\
10 & 0.29160 & 0.26476 & -0.40790 \\
11 & 0.29074 & 0.31604 & -0.21267 \\
12 & 0.29311 & 0.41632 & 0.22811 \\
13 & 0.29274 & 0.36693 & 0.42423 \\
\hline
 \end{tabular}
\end{table}

\begin{figure*}
	\includegraphics[width=2\columnwidth]{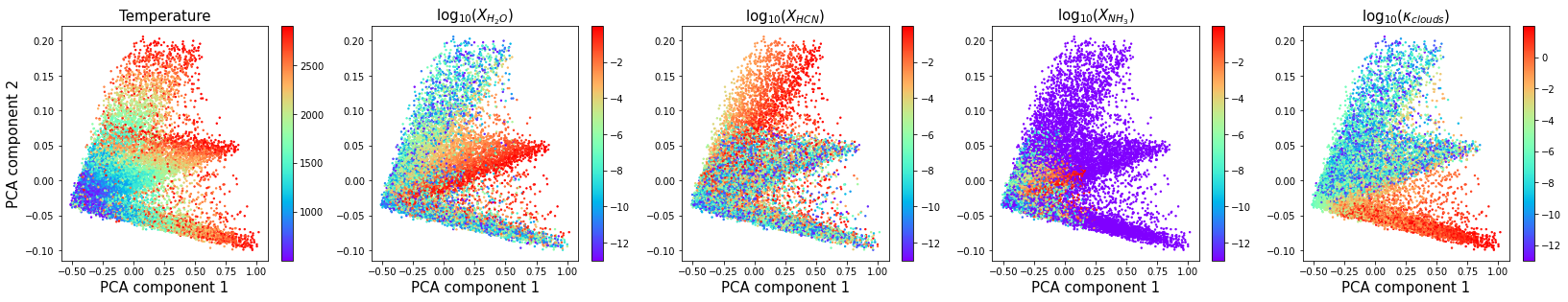}
	\includegraphics[width=2\columnwidth]{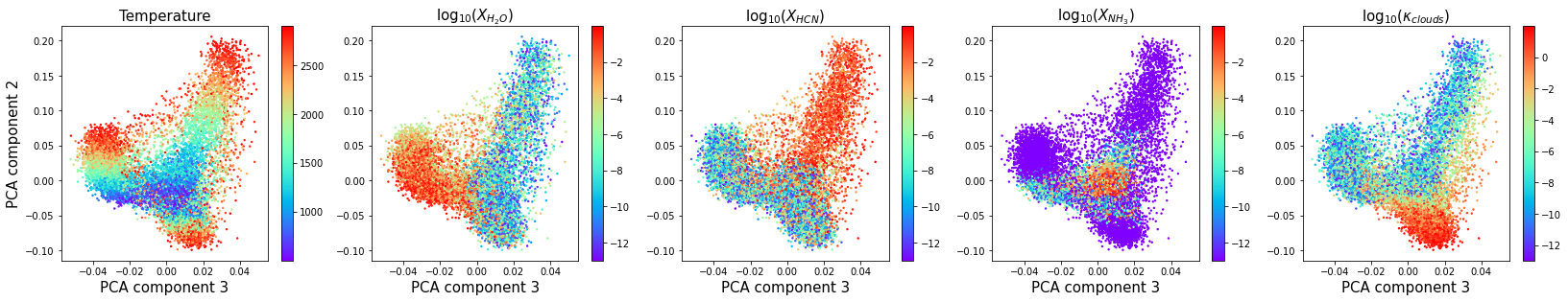}
    \caption{The same as Figure~\ref{fig:pentaptych}, but plotted in the plane of the first and second PCA components (top row) or the plane of the second and third PCA components (bottom row).}
    \label{fig:PCAplane}
\end{figure*}

The effectiveness of the dimensionality reduction achieved through the PCA is illustrated in Figure~\ref{fig:PCAplane}, which is the exact analogue of Figure~\ref{fig:pentaptych}, only this time plotted in the plane of the first and second PCA components (plots in the top row) or in the plane of the second and third PCA components (plots in the bottom row). The plots in the top row are visually very similar to those in Figure~\ref{fig:pentaptych} --- this is not all that surprising, given the correlations exhibited in Figure~\ref{fig:PCAvsStats}. The plots in the bottom row are much more interesting, since the four distinct clusters (high H${}_2$O, high HCN, high ammonia and thick clouds) appear to be well separated. Taken together, the two rows suggest that a three-dimensional representation would be able to capture nicely the existing structure in this data. We shall return to this point in Section~\ref{sec:interpretation} below.

\section{Manifold Learning}
\label{sec:manifold}

\begin{figure*}
	\includegraphics[width=2\columnwidth]{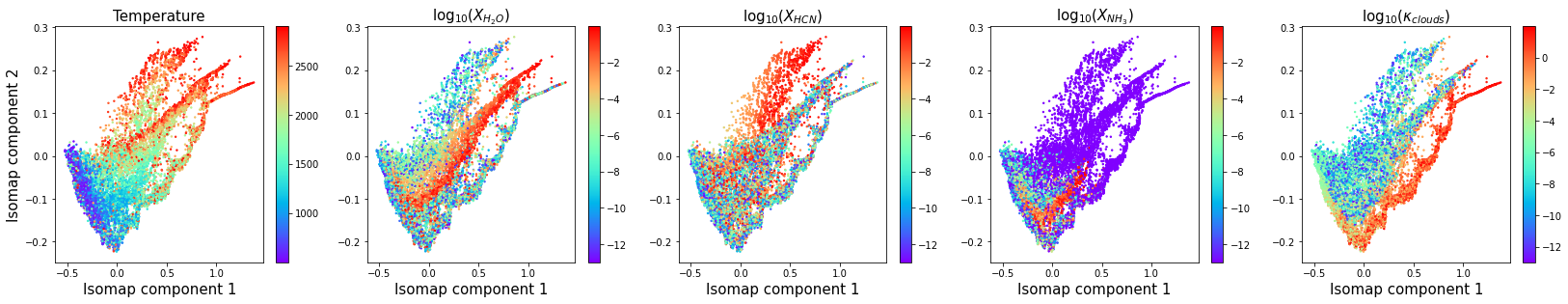}
    \caption{The same as Figure~\ref{fig:pentaptych}, but plotted in the plane of the first two Isomap components.}
    \label{fig:Isomapplane}
\end{figure*}

While PCA is flexible, fast and easily interpretable, it may not perform as well when there are non-linear relationships in the data. For this reason, we also tried several alternative popular methods for dimensionality reduction available in {\tt scikit-learn} \citep{Hastie2001statisticallearning}: Locally Linear Embedding (LLE), Isometric Feature Mapping (ISOMAP), Multi-Dimensional Scaling (MDS), Local Tangent Space Alignment (LTSA), Hessian LLE, t-Distributed Stochastic Neighbor Embedding (t-SNE), and even a simple autoencoder with a two-dimensional latent space and no hidden layers. In each case, we projected down to two degrees of freedom, with mixed results. Some methods failed to unfold to a genuine two-dimensional manifold, and the result was a one-dimensional feature. Others were more successful, and their results were in qualitative agreement with those in Figures~\ref{fig:pentaptych} and \ref{fig:PCAplane}. Figure~\ref{fig:Isomapplane} shows one representative such example, namely the Isomap projection, which does identify the branches corresponding to different constituents. While these manifold learning methods are suitable for very high-dimensional non-linear data, they are less interpretable than PCA and therefore we shall not consider them any further.

\section{Identifying Groups of Similar Spectra}
\label{sec:grouping}

\subsection{Clustering}
\label{sec:clustering}

Clustering is another classic unsupervised task. In the context of exoplanet transmission spectroscopy, \cite{Hayes2020} used clustering to obtain informed priors for the radiative transfer retrieval model, by applying the K-means clustering algorithm to segregate a library of synthetic spectra into discrete classes. The main motivation was to speed up the retrievals by optimizing the initial parameter guess. This motivation governs the choice of the number of classes to be used in the procedure --- while in theory the more classes the better, in practice the optimal number of classes depends on the noise level of the data, which may cause misclassification. The study of \cite{Hayes2020} settled on 30 classes.

Our analysis in Section~\ref{sec:statistics} revealed several distinct groups (branches) in the data which correspond to separate classes of atmospheres. Therefore our goal with the clustering technique is not to find good initial guesses but rather identify the individual branches, which are the interesting regions in terms of physics and chemistry. From our point of view, the number of classes is fixed not by the resolution or noise level, but by the number of interesting physics and/or chemistry regimes, which in our case is related to the number of gas components present in the atmosphere.

\begin{figure}
\begin{center}
	\includegraphics[width=0.48\columnwidth]{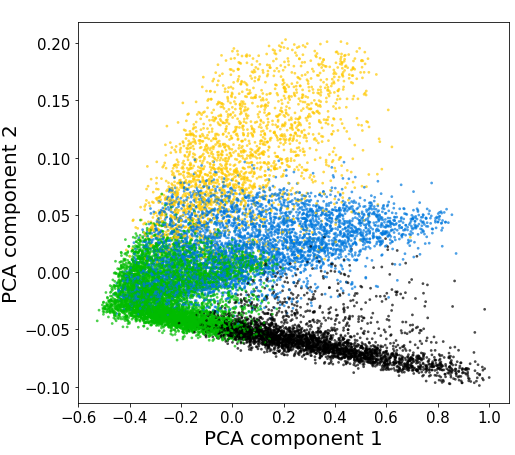}
	\includegraphics[width=0.48\columnwidth]{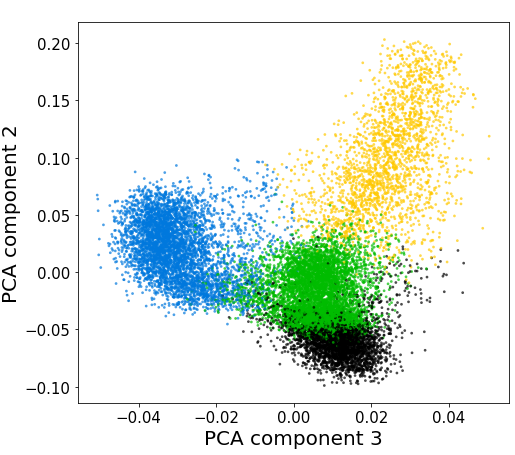}
\end{center}
    \caption{Results from K-means clustering with 4 clusters in the two PCA planes of Figure~\ref{fig:PCAplane}. Note that the clustering was performed in the full 13-dimensional space of standardized PCA components, and the results are then plotted in two dimensions only for visualization purposes. The four clusters found by the algorithm are color-coded with blue, green, yellow and black.}
    \label{fig:clustering}
\end{figure}

We use the K-means clustering tool in {\tt scikit-learn} to analyze the benchmark data set. After standardizing the PCA components, we perform the clustering in the full 13-dimensional PCA space. Figure~\ref{fig:clustering} shows our results for 4 clusters, plotted in the same two PCA planes as in Figure~\ref{fig:PCAplane}: the first and second PCA components in the left panel and the second and third PCA components in the right panel. Upon comparison with Figure~\ref{fig:PCAplane}, we see that the K-means algorithm correctly identified the four interesting regions discussed earlier: the cloud branch (black points), the water branch (blue points), the HCN branch (yellow points) and the ammonia branch (green points). Note that while the points in Figure~\ref{fig:PCAplane} were color-coded using the target information (which is, of course, unavailable in practice), the result in Figure~\ref{fig:clustering} is truly unsupervised, since the classification was done with no prior knowledge of the temperature and composition of the atmosphere!

The success of the unsupervised clustering procedure depicted in Figure~\ref{fig:clustering} offers a promising avenue for future studies in this direction. For example, while the K-means clustering algorithm requires the number of clusters to be specified beforehand, there are alternative clustering algorithms which can hone in on the optimal number of clusters by themselves. Alternatively, one can investigate the optimal number of clusters within K-means itself, e.g., by silhouette analysis. The main lesson from our exercise is that the clustering of the spectral data carries important information about the presence or absence of certain atmospheric constituents. A newly observed spectrum can be immediately associated with one of these physics-motivated clusters, leading to the quick classification of the planet according to its atmospheric chemical composition.

\subsection{Anomaly detection}
\label{sec:anomaly}

Anomaly and outlier detection is another common unsupervised task, where the objective is first to learn what ``normal" data looks like, and then use that to identify unusual instances (observations). Several types of clustering algorithms can be used for this purpose as well, since unusual observations will appear as outliers and will not be easily clustered with the others. The detection of an anomalous transit spectrum could have several important implications:
\begin{itemize}
    \item {\em Discovery of an unusual exoplanet.} This is perhaps the most exciting option, which could indicate the presence of unexpected exotic chemistry, and may even contain clues to extraterrestrial life.
    \item {\em Inadequate data or simulations.} Perhaps not so exciting, but equally important, outcome could be that there are deficiencies with the models used to generate the synthetic database and define the ``normal" categories. The specific reasons could range from missing physics or chemistry to inadequate theoretical approximations or sampling ranges for the relevant parameters.
    \item {\em Instrumental glitches and calibration errors.} Finally, there could be a problem with the observation itself, due to the instrument not functioning properly or issues with the calibration \citep{Azari2020, Azari2021}.
\end{itemize}

\section{Interpretation of the Results}
\label{sec:interpretation}

The analysis in Section~\ref{sec:PCA} revealed that the largest information gain is from including the first three PCA components. Up to now, we have been visualizing the data by projecting on the corresponding two-dimensional PCA planes. However, the disadvantage of this approach is that in doing so, we are losing the higher-dimensional (3D) correlations. This motivates us to extend the previous discussion to three-dimensions, which will better visualize the existing structures and correlations in the data.

\begin{figure*}
\begin{center}
	\includegraphics[width=0.85\columnwidth]{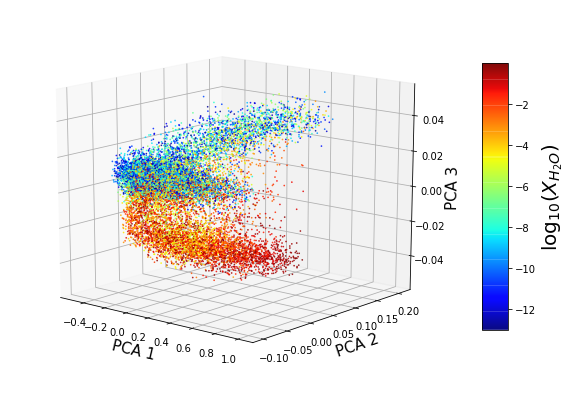}
	\includegraphics[width=0.85\columnwidth]{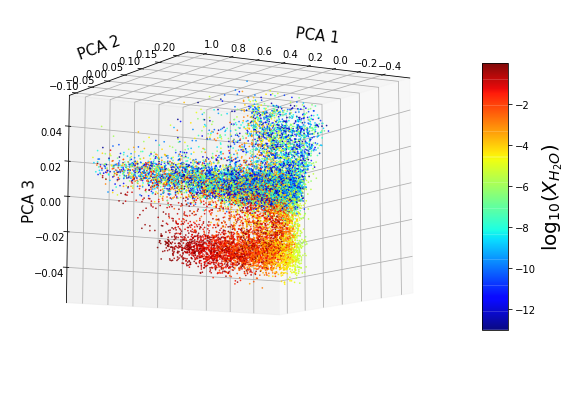}
	\includegraphics[width=0.85\columnwidth]{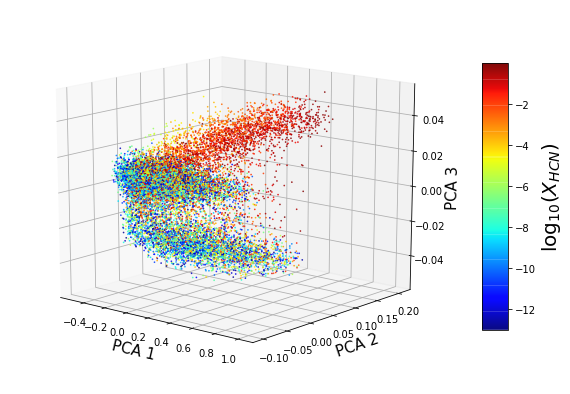}
	\includegraphics[width=0.85\columnwidth]{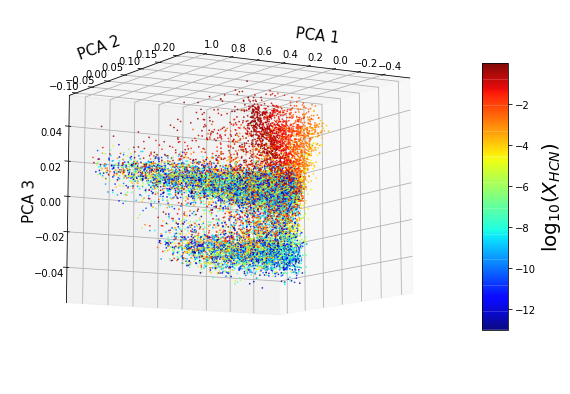}
	\includegraphics[width=0.85\columnwidth]{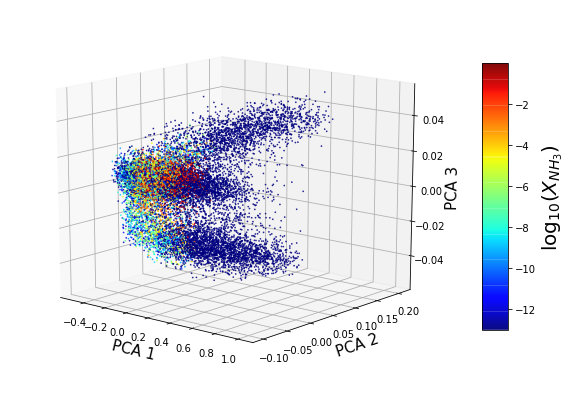}
	\includegraphics[width=0.85\columnwidth]{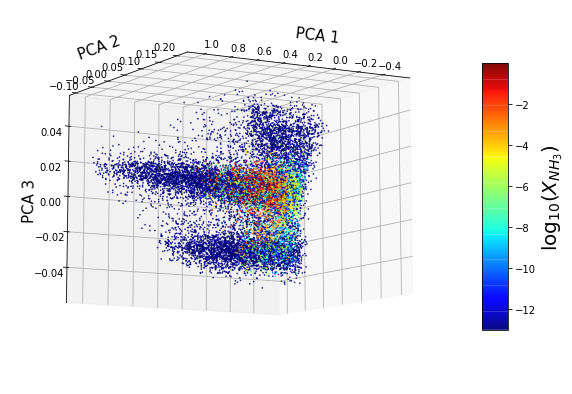}
	\includegraphics[width=0.85\columnwidth]{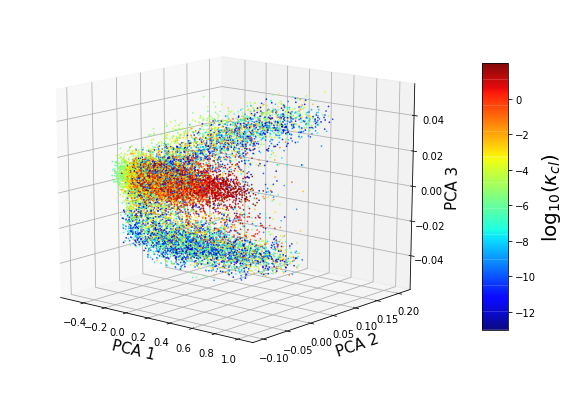}
	\includegraphics[width=0.85\columnwidth]{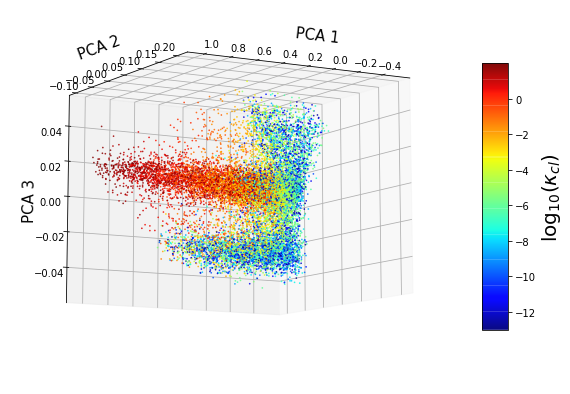}
\end{center}
    \caption{A three-dimensional visualization of the transit spectra in the benchmark database in the 3D space of the first three PCA components. In each panel, the points are color-coded according to one of the input variables: $X_{\rm H_2O}$ (first row), $X_{\rm HCN}$  (second row), $X_{\rm NH_3}$  (third row) and $\kappa_{cl}$  (last row). In the left (right) column the viewpoint is chosen above (below) the PCA1-PCA2 plane.}
    \label{fig:3dplot}
\end{figure*}

Figure~\ref{fig:3dplot} provides a three-dimensional rendering of the transit spectra in the benchmark data set in the space of the first three PCA components. In the left (right) column of plots the viewpoint has been suitably chosen above (below) the PCA1-PCA2 plane to make it easy to see all branches. To guide the eye, in each panel the points are color-coded according to one of the input variables: $X_{\rm H_2O}$ (first row), $X_{\rm HCN}$  (second row), $X_{\rm NH_3}$  (third row) and $\kappa_{cl}$  (last row). We emphasize again that the resolved three-dimensional structure in the PCA space depicted in Figure~\ref{fig:3dplot} was created entirely by analyzing the transit spectra themselves, {\em without any prior knowledge} of the composition or temperature of the atmosphere, i.e., in a true unsupervised manner. This motivates the use of higher-dimensional visualizations of the data in PCA space in discovery mode to characterize the chemical composition of exoplanet atmospheres.
    
\begin{figure}
\begin{center}
	\includegraphics[width=0.8\columnwidth]{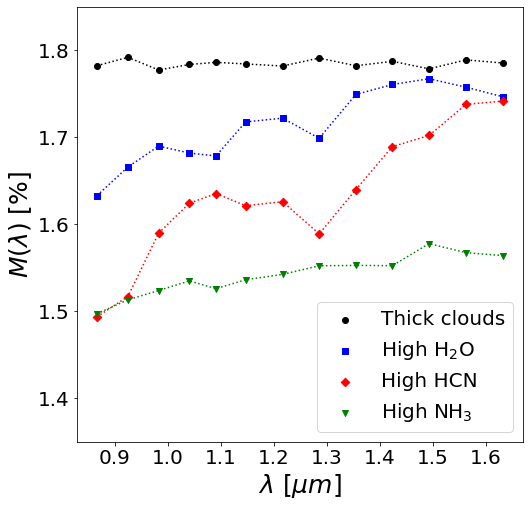}
\end{center}
    \caption{Four transit spectra from the very tip of each of the four different branches discussed in the text: thick cloud (black circles), high H$_2$O (blue squares), high HCN (red diamonds) and high NH$_3$ (green triangles). The input parameters for each spectrum are listed in Table~\ref{tab:inputs}.  }
    \label{fig:spectra}
\end{figure}

In conclusion of this section, for the benefit of the reader, Figure~\ref{fig:spectra} shows four sample spectra corresponding to each of the four different chemical regimes behind the individual branches in Figure~\ref{fig:3dplot} (recall that those branches were also successfully identified by the clustering analysis of Section~\ref{sec:clustering}). The input parameters for each spectrum are listed in Table~\ref{tab:inputs}, together with the corresponding average $\mu(M)$ and standard deviation $\sigma(M)$ for each spectrum. We see that an atmosphere with thick clouds (black circles) has a relatively flat spectrum, while a HCN-rich atmosphere (red diamonds) exhibits a relatively steep slope in the spectrum, with visibly pronounced absorption features. Water-rich atmospheres (blue squares) and ammonia-rich atmospheres (green triangles) fall in between those two extremes. 

\begin{table}
 \caption{The input physical parameters behind the four illustrative spectra shown in Figure~\ref{fig:spectra}. The last two rows show the corresponding average $\mu(M)$ and standard deviation $\sigma(M)$ for each spectrum. }
 \label{tab:inputs}
 \begin{tabular}{lrrrr}
  \hline
       & High H$_2$O & High HCN & High NH$_3$ & Thick clouds \\
  \hline
%
%
%
%
%
%
%
  $T\ [{\rm K}]$            &2828.84 & 2834.14 & 1474.93 & 2884.70\\
  $\log_{10}(X_{\rm H_2O})$ &  -0.47 &  -12.48 &  -11.59 & -12.97  \\
  $\log_{10}(X_{\rm HCN})$  & -12.92 &   -0.49 &  -10.05 & -12.77  \\
  $\log_{10}(X_{\rm NH_3})$ & -13.00 &  -13.00 &   -0.47 & -13.00 \\
  $\log_{10}(\kappa_{cl})$  & -12.36 &  -12.24 &  -11.97 & 1.89     \\
  \hline
$\mu(M)$    & 1.713 & 1.631 & 1.542 & 1.785 \\
$\sigma(M)$ & 0.040 & 0.072 & 0.022 & 0.004 \\  
\hline
 \end{tabular}
\end{table}

Figure~\ref{fig:spectra} illustrates the difficulty in interpreting the observed spectra. Note that the sample spectra shown in the figure are extreme members of the corresponding branches - they all have a very large concentration of a single absorbent, and high temperatures. In more typical situations, the spectra would appear much more similar, complicating the unique interpretation. In contrast, the PCA representation puts each of those spectra squarely on a single branch, thus quickly identifying the correct chemistry. 

\section{Conclusions and Outlook}
\label{sec:conclusions}

The vast majority of data in the world (including the data from transit spectroscopy) is unlabelled.\footnote{At NIPS2016 Yann LeCun famously quipped: ``If intelligence was a cake, unsupervised learning would be the cake, supervised learning would be the icing on the cake, and reinforcement learning would be the cherry on the cake" \cite{geron2017}.} Therefore, in this paper we focused on {\em unsupervised} techniques for analyzing spectral data from transiting exoplanets. We demonstrated methods for i) cleaning and validating the data, ii) initial exploratory data analysis based on summary statistics (estimates of location and variability), iii) exploring and quantifying the existing correlations in the data, iv) pre-processing, standardizing and linearly transforming the data (PCA), v) dimensionality reduction and manifold learning, vi) clustering and anomaly detection, vii) visualization and interpretation of the data. None of these techniques requires any knowledge of the underlying physics and chemistry of the observed atmospheres, and therefore they can serve as useful pre-processing steps for a detailed inversion analysis. In that sense, we view the methodology outlined in this paper as a blueprint for unsupervised analyses and/or preprocessing of transit spectral data which can be implemented as part of inversion pipelines for large exoplanet transit surveys.

We showed that there is a high degree of correlation in the spectral data, which calls for suitable low-dimensional representations. We explored a number of different techniques for such dimensionality reduction and identified several good options, e.g. using summary statistics, PCA decomposition, etc. By analyzing the benchmark data set of \cite{Marquez2018}, we uncovered interesting structures in the data, namely, individual branches corresponding to different chemical regimes. We showed that our two-dimensional representations typically work well to describe those features, but a three-dimensional representation captures the correlations even better. 

Specific results and lessons from our analysis are summarized below:
\begin{itemize}
\item The two-dimensional $(\mu,\sigma)$ representation illustrated in Figure~\ref{fig:pentaptych} is the simplest, fastest and most natural way to quickly characterize the chemical class of the atmosphere of the planet. At the same time, candidate atmospheres which end up in the bulk of the distribution (at low temperatures) are difficult to distinguish.
\item The three-dimensional representation in terms of PCA components illustrated in Figure~\ref{fig:3dplot} nicely resolves the individual classes of atmospheric compositions.
\item We showed that a simple clustering algorithm was able to successfully identify the branches discovered in the 3D PCA space. Such clustering analyses can be used to 
determine (or at least constrain) the number of individual gas components impacting the spectrum. 
\item Transit spectroscopy (or any spectroscopic study in general) deals with high-dimensional, highly-correlated data. PCA and other factorization methods allow to recast the data in a low-dimensional space with minimal loss of information.
\item We provided an intuitive understanding of the PCA decomposition by pointing out that the first PCA component is perfectly correlated with the spectral average while the second PCA component reflects the slope of the function $M(\lambda)$.
\end{itemize}

In this paper we used a toy data set where many parameters were fixed, which means that the data refers to a given single planet. Nevertheless the idea can be suitably generalized, e.g., reformulating in terms of the dimensionless parametrization from \cite{Matchev2021analytical}. This also highlights the need to have much more general data sets available for theoretical studies. The data sets need to be expanded in two separate ways: i) vary the fixed parameters in the current benchmark data set and ii) include potential absorption of other species beyond the four considered here.

\section*{Acknowledgements}
We would like to thank K. Heng for valuable and insightful discussions.
This work was supported in part by the United States Department of Energy under Grant No. DESC0022148.

\section*{Data Availability}

The data underlying this article are described in \cite{Marquez2018} and publicly available in GitHub at \url{https://github.com/exoclime/HELA}.



\bibliographystyle{mnras}
\bibliography{example} 






\bsp	
\label{lastpage}
\end{document}